\documentclass[aps,prb,twocolumn,groupedaddress,showpacs,amsmath,amssymb]{revtex4}

\usepackage[dvips]{graphicx}
\usepackage{color}

\begin{document}

\title{Anomalous suppression of the superfluid density in the
Cu$_x$Bi$_2$Se$_3$ superconductor upon progressive Cu intercalation}

\author{M.~Kriener, Kouji~Segawa, Satoshi~Sasaki, and Yoichi~Ando} 
\affiliation{Institute of Scientific and Industrial Research, Ibaraki,
Osaka University, Osaka 567-0047, Japan}

\date{\today}

\begin{abstract}
Cu$_x$Bi$_2$Se$_3$ was recently found to be likely the first example of a
time-reversal-invariant topological superconductor accompanied by
helical Majorana fermions on the surface. Here we present that
progressive Cu intercalation into this system introduces significant
disorder and leads to an anomalous suppression of the superfluid density
which was obtained from the measurements of the lower critical field. At
the same time, the transition temperature $T_{\rm c}$ is only moderately
suppressed, which agrees with a recent prediction for the
impurity effect in this class of topological superconductors bearing
strong spin-orbit coupling. Those unusual disorder effects give 
support to the possible odd-parity pairing state in Cu$_x$Bi$_2$Se$_3$.
\end{abstract}


%74.20.Mn Nonconventional mechanisms
%74.20.Rp Pairing symmetries (other than s-wave)
%74.25.Bt Thermodynamic properties
%74.62.Dh Effects of crystal defects, doping and substitution

\maketitle

%Introduction

A topological superconductor (TSC) is a superconducting analog of
topological insulators (TIs) and is characterized by a nontrivial
topological structure of the Hilbert space, which is specified by
nontrivial $Z$ or $Z_2$ indices.
\cite{fu08a,schnyder08a,qi09b,qi10b,linder10b,sato10a,qi11a} Its
hallmark signature is the appearance of surface Majorana fermions, which
are their own antiparticles and are of fundamental intellectual
interest. \cite{wilczek} Recently, it was theoretically predicted
\cite{fu10a} and experimentally confirmed \cite{sasaki11a} that a
superconducting doped TI material Cu$_x$Bi$_2$Se$_3$ is likely the first
concrete example of a time-reversal-invariant TSC. Its parent TI
material Bi$_2$Se$_3$ consists of basic crystallographic units of
Se--Bi--Se--Bi--Se quintuple layers, which are weakly bonded by the van
der Waals force. Upon intercalation of Cu into the van der Waals gap,
superconductivity appears below the critical temperature $T_{\rm c}$ of
up to $\sim 3.8$ K. \cite{hor10a} This is high for its low charge
carrier concentration of only $\sim 10^{20}$ cm$^{-3}$. \cite{hor10a} As
a ``superconducting topological insulator", this material has attracted
a great deal of interest.
\cite{fu10a,wray10a,das11,hao11a,ishida11a,kirzhner11a,kriener11a,
kriener11b, wray11a,bay12a,beenakker12a,hsieh12a,michaeli12a,tanaka12a,
ytanaka12a, yamakage12a} 

Unfortunately, Cu$_x$Bi$_2$Se$_3$ has a materials problem in that samples
with a large superconducting volume fraction are difficult to obtain
with the usual melt-growth method, \cite{hor10a,wray10a} which has hindered
detailed studies of the superconducting properties of this material.
However, this problem has  been ameliorated recently by the development
of an electrochemical synthesis technique \cite{kriener11b} which allowed
the synthesis of superconducting samples with the shielding fraction
exceeding 50\%. \cite{kriener11a} Such an improvement made it possible
to perform point-contact spectroscopy on a cleaved surface of
Cu$_x$Bi$_2$Se$_3$ to find a signature of the Andreev bound state in the
form of a pronounced zero-bias conductance peak, \cite{sasaki11a} which
gives evidence for unconventional superconductivity.
\cite{tanaka95a,kashiwaya00a} Knowing that the symmetry of this material
\cite{zhang09a,liu10b} allows only four types of superconducting gap
functions \cite{fu10a} and that all possible unconventional states are
topological, \cite{sasaki11a} it was possible to conclude that
Cu$_x$Bi$_2$Se$_3$ is most likely a TSC. \cite{sasaki11a}

Although the point-contact spectroscopy elucidated the possible TSC nature of
Cu$_x$Bi$_2$Se$_3$, the electron mean free path $\ell$ in the
superconducting samples of this material is comparable to the coherence
length $\xi_0$; \cite{kriener11a} according to the common belief,
\cite{balian63a,larkin65a} the odd-parity pairing should be strongly
suppressed by impurity scattering in such a situation.
\cite{mackenzie03a,maeno12a} In this context, a recent theory by
Michaeli and Fu addressed this issue \cite{michaeli12a} and showed that
odd-parity superconductivity in strongly spin-orbit coupled
semiconductors such as Cu$_x$Bi$_2$Se$_3$ are much more robust against the
pair-breaking effect induced by impurity scattering than in more
ordinary odd-parity superconductors. Therefore, thanks to the role of
spin-orbit coupling, $T_{\rm c}$ of Cu$_x$Bi$_2$Se$_3$ is expected to be
rather insensitive to nonmagnetic impurities, which is similar to
conventional superconductors. \cite{anderson59b} 

In this Rapid Communication, we address the issue of disorder effects in
Cu$_x$Bi$_2$Se$_3$. Through our systematic studies of the effects of Cu
intercalation in this system, it turned out that increasing the Cu
content beyond $x \sim$ 0.3 in the superconducting regime does not
increase $T_{\rm c}$ or the carrier concentration, but its main effect is to
enhance the residual resistivity $\rho_0$. This suggests that one can
consider the Cu content $x$ to be a parameter to control the disorder
while keeping other fundamental parameters virtually unchanged. By
looking at the data from this perspective, the $x$ dependence of the
superfluid density obtained from the lower critical field shows an
unusual disorder dependence that is distinct from that in conventional
BCS superconductors, which gives support to unconventional
pairing. In addition, we show that the $x$ dependence of $T_{\rm c}$ is
essentially a reflection of the disorder effect and is consistent with
the particular odd-parity paring state that is supposed to be realized
in Cu$_x$Bi$_2$Se$_3$.

%Experiment: Samples and Methods

\begin{figure}[t]
\centering
\includegraphics[width=8.5cm,clip]{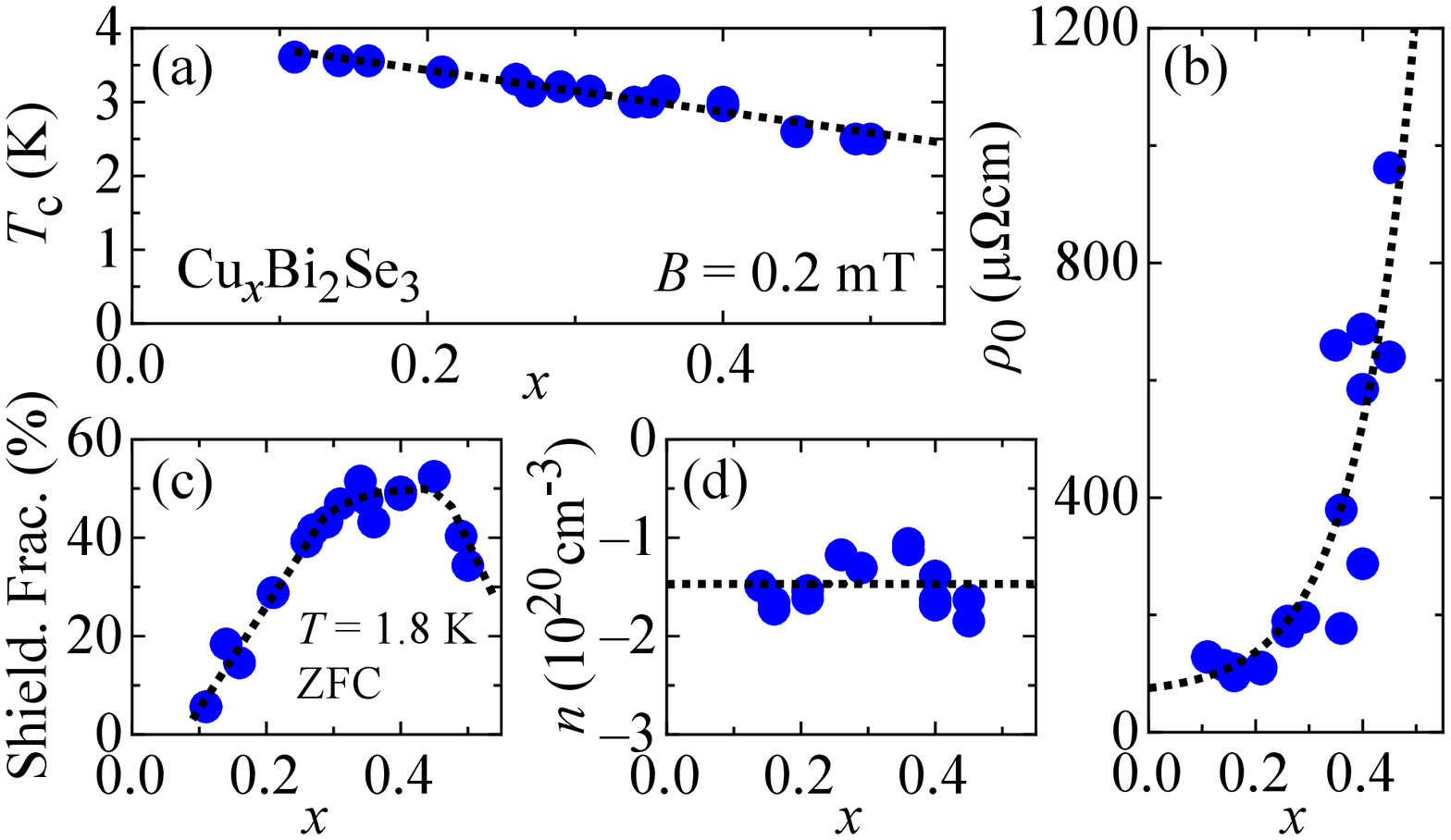}
\caption{(Color online) Cu content $x$ dependences of (a) critical 
temperature $T_{\rm c}$, (b) residual resistivity $\rho_0$, (c) 
superconducting shielding fraction, and (d) normal-state carrier 
density $n$. For comparison, the carrier density
in pristine Bi$_2$Se$_3$ ($\lesssim 10^{19}$ cm$^{-3}$) is shown 
with a diamond. 
Note that $n$ is essentially $x$ independent and 
remains low at $\sim 1.5\times 10^{20}$ cm$^{-3}$. 
The dotted lines are guides to the eyes.}
\label{fig1}
\end{figure}

Cu$_x$Bi$_2$Se$_3$ single crystals of slab-like geometry with various Cu
contents $0.11 \leq x \leq 0.50$ were prepared by the electrochemical
technique described earlier. \cite{kriener11b} The typical sample size
was $4 \times 2.5 \times 0.3$ mm$^3$. The magnetic field dependence of
the magnetization, $M(B)$, were measured with a commercial superconducting 
quantum interference device (SQUID) 
magnetometer (Quantum Design MPMS) with particular attention being paid to the
low field regime. \cite{MTcomment} Roughly half of the samples were also
characterized by transport measurements by a standard six-probe method.
Figure~\ref{fig1} summarizes the $x$ dependences of $T_{\rm c}$,
$\rho_0$ (defined as $\rho$ at $T$ = 5 K), the superconducting shielding
fraction at $T$ = 1.8 K, and the charge carrier concentration $n$
(determined from the Hall coefficient at 5 K). Most notably, $\rho_0$
strongly increases for $x>0.3$ and $n$ is basically independent of $x$
at $n \simeq 1.5 \times 10^{20}$ cm$^{-3}$.\cite{note_n} 
 
%Figure 1

Before presenting magnetic properties, we define and summarize important
parameters. The layered structure of Cu$_x$Bi$_2$Se$_3$ leads to
anisotropies in the superconducting parameters, and we denote the lower
and upper critical fields for magnetic fields parallel and perpendicular
to the crystallographic $ab$ planes as $B_{{\rm c1},ab}$, $B_{{\rm
c1},c}$, $B_{{\rm c2},ab}$, and $B_{{\rm c2},c}$, respectively. Also, the
penetration depths and the coherence lengths along the in-plane and
out-of-plane directions are denoted as $\lambda_{ab}$, $\lambda_{c}$,
$\xi_{ab}$, and $\xi_c$, respectively. The anisotropic Ginzburg-Landau
parameters are defined as
$\kappa_{ab}=\sqrt{\lambda_{ab}\lambda_c/(\xi_{ab}\xi_c)}$ and
$\kappa_{c}=\lambda_{ab}/\xi_{ab}$. \cite{klemm80a,kogan81a,clem89a} The
upper critical fields are related to the coherence lengths via $B_{{\rm
c2},ab} = \Phi_0/2\pi\xi_{ab}\xi_c$ and $B_{{\rm c2},c} =
\Phi_0/2\pi\xi_{ab}^2$ with the flux quantum $\Phi_0$. In the
Ginzburg-Landau theory, $B_{\rm c1}$ is related to the vortex line
energy $E$ via $B_{\rm c1}=4\pi \mu_0 E/\Phi_0$; \cite{parks69b} for
extremely type-II superconductors with $\kappa \gg 1$, one obtains
$E\approx [\Phi_0^2/(4\pi \lambda)^2]\ln \kappa$. However, to take into
account the vortex core energy, the $\ln \kappa$ term has to be
corrected by adding 0.5 \cite{abrikosov57a,hu72a,liang05a}, and the
formula for $B_{{\rm c1},ab}$ becomes
\begin{equation}\label{eq1}
B_{{\rm c1}, 
ab}=\frac{\Phi_0}{4\pi}[\ln(\kappa_{ab})+0.5]\frac{1}{\lambda_{ab}\lambda_c}.
\end{equation} 
Hence, to calculate the Ginzburg-Landau parameter $\kappa_{ab}$, we use 
$B_{{\rm c1},ab}/B_{{\rm c2},ab} = (\ln\kappa_{ab}+0.5)/2\kappa_{ab}^2$. 
The anisotropy factor is defined as 
$\gamma\equiv B_{{\rm c2},ab}/B_{{\rm c2},c}= 
B_{{\rm c1},c}/B_{{\rm c1},ab}=\lambda_{c}/\lambda_{ab}$ and
the penetration 
depths are determined by solving Eq.~\ref{eq1} for $\lambda_{ab}$ by using 
$\lambda_c = \gamma \lambda_{ab}$. For the following discussion, we define the 
averaged penetration depth $\lambda_{\rm av}=\sqrt[3]{\lambda_{ab}^2 
\lambda_c}$, which allows the calculation of the superfluid density via 
$n_{s} =m^*/(\mu_0 e^2 \lambda_{\rm av}^2)$ with the effective mass 
$m^*$ assumed to be $x$ independent. \cite{cpcomment}

%Results

%Figure 2
\begin{figure}[t]
\centering
\includegraphics[width=8.5cm,clip]{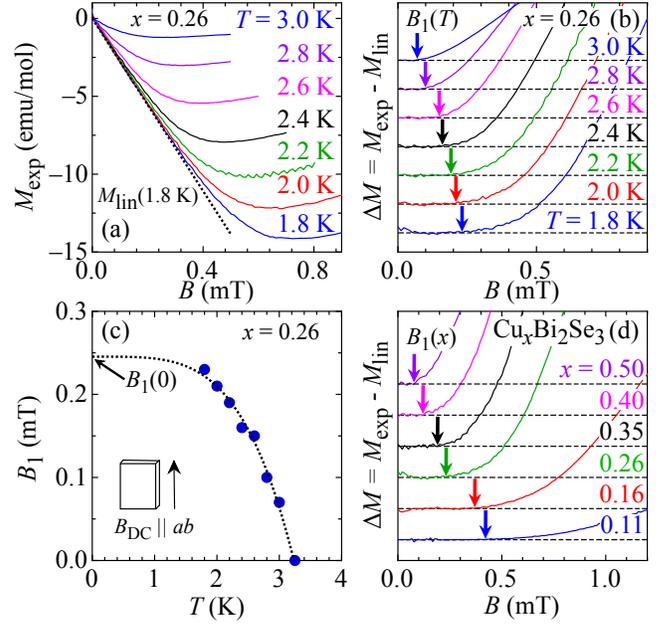}
\caption{(Color online) (a) Initial low-field $M_{\rm exp}$ vs $B$
curves for a sample with $x$ = 0.26 at various temperatures. (b) Reduced
magnetization $\Delta M$ after subtracting the initial linear Meissner
contribution $M_{\rm lin}$. The deviation points marked by arrows
indicate $B_1$ at each temperature. (c) Plot of $B_1$ vs $T$ for $x$ =
0.26 together with a fit to the data. 
(d) Plots of $\Delta M$ vs $B$ at 1.8 K for various $x$.} 
\label{fig2}
\end{figure}

Figure 2 describes how $B_{{\rm c1},ab}$ is determined from the magnetization
data $M_{\rm exp}$, which is essentially the same as was done in Ref.
\onlinecite{kriener11a}. Figure 2(a) shows $M_{\rm exp}(B)$ curves for a
sample with $x$ = 0.26 at various temperatures between 1.8 and 3.0 K;
the $T_{\rm c}$ of this sample was $\sim$3.2 K. The dashed line $M_{\rm
lin}$ is a fit to the low-field magnetization at 1.8~K, representing the
initial Meissner screening. Determining such a linear part for each
temperature and subtracting it from the $M_{\rm exp}(B)$ data yields
$\Delta M = M_{\rm exp}-M_{\rm lin}$ which is shown in Fig. 2(b) (the
data are shifted for clarity). The arrows mark the last field above
which $\Delta M$ shows an obvious deviation from zero (i.e. $\Delta M$
becomes $\gtrsim 0.02$), signaling the entry of vortices and defining
the flux entry field $B_1(T)$.
These data points are plotted as $B_1$ vs
$T$ in Fig. 2(c) and are fitted with the empirical formula $B_{\rm 1}(T)
= B_{\rm 1}(0)[1-(T/T_{\rm c})^4]$. \cite{empcomment} Sometimes the
$B_{\rm 1}(T)$ data scatter around the fitting line, which leads to a
sample-dependent error bar on $B_1(0)$. To show the trend of how $B_1$
changes with the Cu concentration $x$, the 1.8-K magnetization data for
various $x$ are shown in Fig. 2(d); this plot clearly suggests that
$B_1(0)$ becomes systematically smaller for larger $x$.

To determine $B_{\rm c1}$ from $B_1$, one should consider the influences
of the demagnetization effect, the surface quality (Bean-Livingston
surface barrier), and the bulk pinning effects. The surface barrier is
not effective in rough surfaces, which is the case in all of our
as-grown samples (see Fig. S4 of Ref. \cite{sasaki11a}) irrespective of
the $x$ values, and the bulk pinning is also extremely weak in
Cu$_x$Bi$_2$Se$_3$ as indicated by magnetic hysteresis data.
\cite{kriener11a,note_pinning} As for the demagnetization effect, albeit
small ($<10$\%) in the present case, we have corrected for it by using
the Brandt's formula for slab-shaped samples with an aspect ratio $b/a$:
\cite{brandt99a} 
$B_{\rm c1}(0)= B_1(0) / \tanh\sqrt{0.36\,b/a}$. 
The obtained $B_{\rm c1}$ for all samples are plotted vs $x$ in Fig. 3(a).
The corresponding $\lambda_{\rm av}$ values are shown in Fig. 3(b); for
calculating $\lambda_{\rm av}$, we need the anisotropy factor $\gamma$
which was obtained from anisotropic $B_{\rm c2}$ determined from the
resistive transitions in magnetic fields applied parallel and
perpendicular to the $ab$ plane [Fig. 3(c)]. The obtained $\gamma$ is
essentially independent of $x$ [Fig. 3(d)], which supports the idea that
the main effect of Cu intercalation beyond $x \sim$ 0.3 is to enhance
the disorder without changing band structure or mobile carrier density.

%Figure 3
\begin{figure}[t]
\centering
\includegraphics[width=8.5cm,clip]{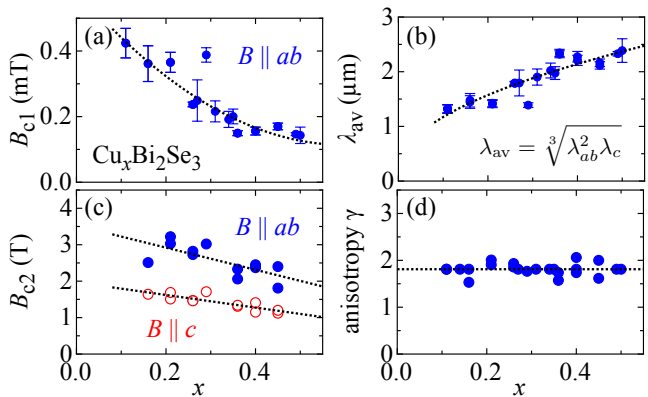}
\caption{(Color online) $x$ dependences of (a) $B_{\rm c1}(0)$ for 
$B \parallel ab$ after the demagnetization correction, 
(b) corresponding penetration depth 
$\lambda_{\rm av}$, (c) $B_{\rm c2}(0)$ for fields parallel and 
perpendicular to the $ab$ planes, and (d) anisotropy factor $\gamma$. 
Dotted lines are guides to the eyes.} 
\label{fig3}
\end{figure}

%Discussion

As already mentioned, the averaged penetration depth $\lambda_{\rm av}$
directly gives the superfluid density $n_s =m^*/(\mu_0 e^2 \lambda_{\rm
av}^2)$. \cite{inhomogeneity} We normalize this value with the
normal-state carrier density $n$, and Fig. 4(a) summarizes the $x$
dependence of $n_s^{\rm exp} \equiv n_s/n$. One can see that $n_s/n$ is
already only 0.3 at $x \simeq$ 0.10 where the superconductivity starts
to appear, and it is further suppressed with increasing $x$. This
behavior is obviously a reflection of strong disorder caused by Cu
intercalation that can be inferred in Fig. 1(b). Since it is known that
disorder causes a reduction in $n_s$ even in conventional BCS
superconductors, \cite{ma85a,balatsky06a} it is prudent to discuss this
behavior quantitatively. 
 
According to Anderson's theorem, \cite{anderson59b} the
superconducting gap $\Delta_0$ and $T_{\rm c}$ of conventional
superconductors are relatively insensitive to small concentrations of
nonmagnetic impurities. However, the superfluid density, which reflects
the ``rigidity" of the electronic system to electromagnetic
perturbations, is affected by disorder in conventional superconductors.
\cite{ma85a,balatsky06a,tallon06a} Indeed, the disorder dependence of
$n_s$ has been studied in Nb and Pb and was found to follow the
theoretical prediction. \cite{desorbo63a,egloff83a} We therefore compare
the disorder dependence of $n_s$ observed in Cu$_x$Bi$_2$Se$_3$ to the
expectation for ordinary BCS superconductors. For such a comparison, one
needs to parametrize disorder, which is usually done by evaluating
$k_{\rm F}\ell$, where $k_{\rm F} = \sqrt[3]{3\pi^2 n}$
is the Fermi wave number \cite{freeleccomment} and $\ell=\hbar k_{\rm
F}/(\rho_0 n e^2)$ is the mean free path. \cite{kFellcomment} 

For a pure BCS superconductor, the penetration depth in the 0-K limit is
given by $\lambda_{\rm L}^2(0)=m^*/(\mu_0 e^2 n)$, because $n_s$ is
equal to $n$ in the clean limit. In the presence of disorder, this
$\lambda_{\rm L}(0)$ in the clean limit is modified to an effective
penetration depth which is evaluated at $T$ = 0 K as $\lambda_{\rm
BCS}(0) =\lambda_{\rm L}(0)\sqrt{1+\xi_0/\ell}>\lambda_{\rm L}(0)$ in
the local limit, \cite{raychaudhuri83a,tinkham96p} where $\xi_0=\hbar
v_{\rm F}/(\pi \Delta_0)$ is the Pippard coherence length for pure
superconductors ($v_{\rm F}=\hbar k_{\rm F}/m^*$ is the Fermi velocity
and $\Delta_0$ is the BCS gap). \cite{delta_comment} From this
$\lambda_{\rm BCS}(0)$ we calculate the superfluid density $n_s^{\rm
BCS}(\ell)$, which gives the disorder-induced suppression of $n_s$ for a
conventional BCS superconductor. 

\begin{figure}[t]
\centering
\includegraphics[width=8.5cm,clip]{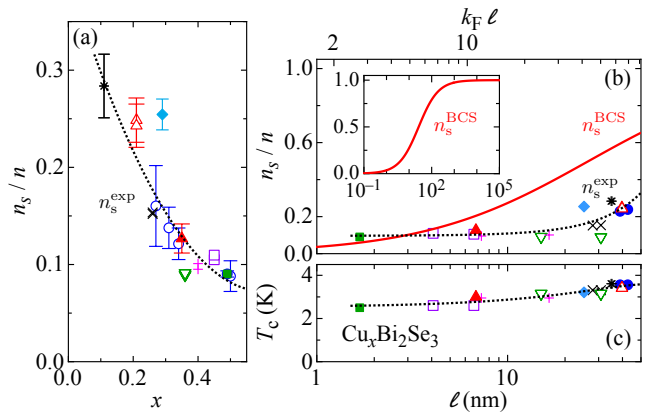}
\caption{(Color online) (a) Normalized superfluid density 
$n_s/n$ vs $x$, with $n$ the normal-state carrier density. 
For symbols without error bars, the estimated errors are 
smaller than the symbol size. (b), (c) Semi-log plots of $n_s/n$ (b) and 
$T_{\rm c}$ (c) vs the mean free path $\ell$. The upper axis gives the corresponding 
$k_{\rm F}\ell$ value. To facilitate comparisons between (a) and (b) [(c)], 
data for different $x$ values are indicated by different symbols. 
The solid line $n_s^{\rm BCS}$ in (b) gives 
the expected disorder-induced suppression of $n_s/n$ for a conventional 
BCS superconductor. The inset shows $n_s^{\rm BCS}\rightarrow 1$ in the 
clean limit $\ell\rightarrow \infty$. The dotted lines in all panels are guides 
to the eyes.}
\label{fig4}
\end{figure}

Figure 4(b) shows the comparison of the $\ell$ dependences of $n_s^{\rm
exp}$ and $n_s^{\rm BCS}$ ($k_{\rm F}\ell$ value is shown in the upper
axis). In this figure, the BCS calculation is shown as a solid line and
the inset shows the saturation of $n_s^{\rm BCS} \rightarrow 1$ in the
clean limit $\ell\rightarrow \infty$. Clearly, $n_s^{\rm exp}$ does not
agree with $n_s^{\rm BCS}$; although both are suppressed with decreasing
$\ell$, the suppression is much stronger in Cu$_x$Bi$_2$Se$_3$ than is
expected for a BCS superconductor. Also, it is useful to compare the
result shown in Fig. 4(b) to that in Fig. 1(b): At $x >$ 0.3, the
residual resistivity starts to increase drastically and $\ell$ becomes
shorter than $\sim$25 nm; however, $n_s^{\rm exp}$ tends to saturate in
this dirtier range of $\ell <$ 25 nm. Moreover, for $\ell <$ 4 nm,
$n_s^{\rm exp}$ intersects the $n_s^{\rm BCS}$ curve. Hence, both the
strong suppression in the intermediate disorder regime and the
saturation tendency in the dirtier regime are anomalous. Such an
anomalous behavior of $n_s/n$ is the main result of this work, and it
naturally points to an unconventional nature of the superconducting
state in Cu$_x$Bi$_2$Se$_3$.

In contrast to the highly anomalous behavior of $n_s^{\rm exp}$, the
modest suppression of $T_{\rm c}$ shown in Fig. 4(c) resembles the
behavior of dirty conventional superconductors. \cite{anderson59b} One
might hasten to conclude that such an ordinary disorder dependence of
$T_{\rm c}$ speaks against the odd-parity pairing, because the common belief
for odd-parity superconductors is that $T_{\rm c}$ is quickly suppressed
with impurity-induced disorder. \cite{balian63a,larkin65a} However, as
we already mentioned above, the particular type of
odd-parity pairing that is considered to be realized in
Cu$_x$Bi$_2$Se$_3$ (Refs. \onlinecite{fu10a,sasaki11a}) belies this
common belief. This point was recently shown by Michaeli and Fu,
\cite{michaeli12a} who analyzed the novel inter-orbital, odd parity
state \cite{fu10a} proposed for Cu$_x$Bi$_2$Se$_3$. The odd-parity
pairing takes place between two $p_z$ orbitals with different parity at
the upper and lower ends of the quintuple layers via attractive
short-range interactions. In such a state, the crucial disorder-induced
pair breaking effect is significantly suppressed as a result of strong
spin-momentum locking. The dephasing rate of the Cooper pairs depends on
the ratio of band mass and chemical potential, $m/\mu$; as this ratio
becomes smaller, the superconductivity becomes more robust. For
Cu$_x$Bi$_2$Se$_3$, this ratio has been estimated \cite{wray11a} to be
$\sim 1/3$ and the calculated $T_{\rm c}$ depends only weakly on the
impurity-induced disorder, \cite{michaeli12a} in qualitative agreement
with Fig. 4(c). Therefore, the observed disorder effect in $T_{\rm c}$ is not
inconsistent with the odd-parity pairing.

% Summary

To summarize, we report an anomalous suppression of the superfluid
density $n_s/n$ probed by the lower critical field as a function of the
Cu content $x$. Since it appears that the main effect of Cu
intercalation beyond $x \sim$ 0.3 is to enhance disorder without
significantly changing band structure or carrier density, our result
reveals the impact of disorder on the superconducting state in
Cu$_x$Bi$_2$Se$_3$. Most strikingly, in the intermediate range of
disorder, $n_s/n$ is much more strongly suppressed than is expected for
a dirty conventional BCS superconductor, while in the strongly
disordered regime $n_s/n$ tends to saturate. In contrast, the occurrence
of superconductivity itself is robust against disorder as indicated by
an only moderate suppression of $T_{\rm c}$ with $x$. The obviously anomalous
behavior in $n_s/n$ points to an unconventional pairing state, and the
ostensibly normal behavior in $T_{\rm c}$ is consistent with the 
theoretically-proposed odd-parity
pairing state with strong spin-momentum locking. Altogether, our result
gives support to the possible odd-parity pairing state in
Cu$_x$Bi$_2$Se$_3$.

% acknowledgement

We thank L. Fu and Y. Tanaka for fruitful discussions, and S. Wada for
technical assistance. This work was supported by JSPS (NEXT Program), 
MEXT (Innovative Area ``Topological Quantum Phenomena"
KAKENHI 22103004), and AFOSR (AOARD 124038).


\begin{thebibliography}{10}
\bibitem{MTcomment}
The magnetization was measured at several temperatures between 1.8 K and
$T_{\rm c}$ of each sample. Before each run, the sample temperature was
stabilized at 5 K ($> T_{\rm c}$) and the magnet was quenched to remove
any remnant field. Then the sample was cooled in zero field to the
desired temperature and the magnetization was measured with increasing
magnetic field.
\bibitem{note_n}
It has been found that $n$ is nearly unchanged among various samples in
the range of $0.1 \le x \le 0.5$ irrespective of the superconducting
volume fraction, and hence it is most reasonable to infer that the local
carrier density is essentially the same in superconducting and
non-superconducting regions.
\bibitem{cpcomment}
The effective mass $m^*$ was estimated for $x=0.29$ from normal-state 
specific-heat data \cite{kriener11a} to be $m^* = 2.6 m_{e}$ ($m_e$ is 
the bare electron mass).
\bibitem{note_pinning} 
We have measured $M(B)$ curves in $\pm$1 T for $x$ = 0.16,
0.29, and 0.40, and found that the pinning is always very weak with the
irreversibility field of around 100 mT which is essentially independent
of $x$.
\bibitem{freeleccomment}
Since the superconducting Cu$_x$Bi$_2$Se$_3$ exhibits only a single ellipsoidal 
Fermi surface around the $\Gamma$ point \cite{wray11a} 
and the electron correlations are expected to be weak due to 
the low-carrier-density nature, the free-electron model is a 
reasonable approximation for calculating the averaged $k_{\rm F}$.
\bibitem{kFellcomment}
Although the disorder increases with $x$, 
$k_{\rm F}\ell$ is still much larger than 1 and hence 
Cu$_{x}$Bi$_2$Se$_3$ is in the weakly disordered regime
\cite{ma85a,balatsky06a}. Also, the ratio $\xi_0/\ell$ is about 1 for 
$x \approx 0.26$ and hence Cu$_{x}$Bi$_2$Se$_3$ is in between the clean and 
dirty limits.
\bibitem{delta_comment}
For our calculation of $\lambda_{\rm BCS}(0)$ expected for BCS
superconductors, we used the BCS gap value 
$\Delta_0= 1.764\,k_{\rm B} T_{\rm c}^0$ with
the clean-limit critical temperature $T_{\rm c}^0$ which we identify
with the maximum $T_{\rm c}$ of 3.8 K in Cu$_x$Bi$_2$Se$_3$.
\end{thebibliography}
\end{document}